# Spintronic Oscillator based on magnetic field feedback

D. Dixit[1], K. Konishi[2], C. V. Tomy[1], Y. Suzuki[2], A. A. Tulapurkar[3]

[1]Department of Physics, Indian Institute of Technology-Bombay, Powai, Mumbai 400076, India

[2]Graduate School of engineering Science, Osaka University, Toyonaka, Osaka 560-8531, Japan

[3]Department of Electrical Engineering and Centre of Excellence in Nanoelectronics, Indian Institute of Technology-Bombay, Powai, Mumbai 400076, India

*We present a circuit design of a spintronic oscillator based on magnetic tunnel junction. In this design, a dc current is passed through a magnetic tunnel junction which is connected to a "feed-back" wire below it. Any fluctuation in the magnetization direction of the free layer of MTJ, drives a fluctuating current through the feed-back wire, which exerts a magnetic field on the free layer. This in turn can amplify the magnetization fluctuations of the free layer. If the dc current passing through the MTJ is more than a critical value, continuous precessing states of the magnetization are possible.*

Development of nano-scale RF oscillators based on the magnetic tunnel junctions is an active area of research. These oscillators are based on the spin-transfer torque effect.[1-3] A dc current, passing through a MTJ, gets polarized by the fixed layer, and exerts a spin-transfer torque (STT) on the free layer. This torque is used to drive the free layer into continuous precessional states.[4,5] Here we present an alternative feedback scheme (not based on the spin-transfer torque effect) to drive a MTJ into spontaneous oscillations. This scheme offers a possibility of better quality factor of the oscillations. Further, STT and this feedback scheme can be combined in the same device for designing oscillators.

The schematic diagram of the oscillator is shown in the Fig.1. The easy direction of the free layer is taken to be along the x-axis, in-plane hard direction along the y-axis and the out-of-plane hard direction along the z-axis. The pinned layer magnetization is taken to be along the y-axis. The MTJ rests on the top of a waveguide which is electrically insulated from the MTJ.[6,7] The top and bottom electrodes of the MTJ are connected through a bias-T to the wave guide (feedback wire) as shown in Fig.1. The feed-back wire is oriented in such a way that a current passing through it creates a magnetic field along the y-axis. As shown in the figure a delay element is also present in the circuit to adjust the phase of the current. The basic working principle of the oscillator is as follows: If a dc current is passed through the MTJ, the thermal fluctuations of the free layer produce oscillating voltage across it due to the magneto-resistance effect. This oscillating voltage drives an oscillating current through the feed-back wire, which in turn creates an oscillating magnetic field along the y-axis. (The dc current passing through MTJ is blocked by the bias-T from flowing through the feed-back wire). By choosing a proper phase of the oscillating current (and thus the ac magnetic field), the magnetization fluctuations of the free layer can be amplified. If the dc current exceeds a critical value, the free layer can undergo spontaneous oscillations.

The magnetization dynamics of the free layer (without the feedback circuit) is given by Landau-Lifshitz-Gilbert (LLG) equation as:[8, 9, 10]

$$\frac{d\hat{m}}{dt} = -\gamma[\hat{m}\times(\overline{H}+\overline{h}_r)] \pm \alpha\gamma\,\hat{m}\times[\hat{m}\times(\overline{H}+\overline{h}_r)] \quad ---(1)$$

where $\hat{m}$ is a unit vector along magnetization, $H$ is the effective magnetic field (including the anisotropy field and external magnetic field) and $\gamma$ is related to the gyromagnetic factor $\gamma_0$ as $\gamma = \gamma_0/(1+\alpha^2)$ where α is the damping factor. The positive sign in the above equation is to be taken for forward direction of time and negative sign for backward direction of time. $h_r$ denotes the random magnetic field due to thermal fluctuations with following statistical properties:

$$\langle h_{r,i}(t) \rangle = 0, \quad \langle h_{r,i}(t) h_{r,j}(s) \rangle = 2D\delta_{ij}\delta(t-s), \quad D = \frac{\alpha}{1+\alpha^2} \frac{k_B T}{\gamma \mu_0 M_S V} \quad ---(2)$$

where < > denotes ensemble average and $M_S$ denotes the magnetization and $V$ denotes the volume of the free layer. (We have neglected STT term in above equation. This is reasonable if the free layer magnetic moment is large). Also here we are interested in proving the existence of spontaneous oscillations without STT term. The effective field can be written as $\bar{H} = -(1/\mu_0 M_s)(\partial E_{mag}/\partial \hat{m})$ where $E_{mag}$ is magnetic energy per unit volume due to Zeeman interaction and anisotropy energy, which can be written as
$E_{mag} = \mu_0 M_s [-H_{ext} m_x + (1/2) H_{//} m_y^2 + (1/2) H_\perp m_z^2]$ Here we have taken the free layer to be in xy plane. The external magnetic field is assumed to along x-axis. $H_{//}$ and $H_\perp$ denote the in-plane and out of plane effective anisotropy fields. Thus the effective magnetic field can be written as $H = H_{ext} \hat{x} - H_{//} m_y \hat{y} - H_\perp m_z \hat{z}$. Due to the random magnetic fields appearing in the LLG equation, the direction of magnetization undergoes fluctuations. We now assume that fluctuations of magnetization are small i.e. $m_x \approx 1, m_y, m_z \ll 1$. We therefore linearize LLG equation and also neglect terms of the form $m_y h_r$ and $m_z h_r$. The resulting LLG equation is written below:

$$\dot{m}_y = -\gamma(H_\perp + H_{ext})m_z \mp \alpha\gamma(H_{//} + H_{ext})m_y + \gamma h_{r,z} \pm \alpha\gamma h_{r,y} \quad (3a)$$
$$\dot{m}_z = \gamma(H_{//} + H_{ext})m_y \mp \alpha\gamma(H_\perp + H_{ext})m_z - \gamma h_{r,y} \pm \alpha\gamma h_{r,z} \quad (3b)$$

We can consider our system to be represented by an ensemble of systems each of which obeys above stochastic differential equation. We are interested in looking at systems in the ensemble whose $m_y$ and $m_z$ match with the given values at say t=0. We now take ensemble average of above equations[10] for these systems to obtain the following equations:

$$\frac{d}{dt}\langle m_y \rangle = -\gamma(H_\perp + H_{ext})\langle m_z \rangle \mp \alpha\gamma(H_{//} + H_{ext})\langle m_y \rangle \quad (4a)$$
$$\frac{d}{dt}\langle m_z \rangle = \gamma(H_{//} + H_{ext})\langle m_y \rangle \mp \alpha\gamma(H_\perp + H_{ext})\langle m_z \rangle \quad (4b)$$

The above equations give the solution for average values of $m_y$ and $m_z$ as $\langle m_y \rangle = \text{Re}[Ae^{-i\omega_0 t} e^{-|t|/\tau_0}]$ and $\langle m_z \rangle = \text{Re}[Be^{-i\omega_0 t} e^{-|t|/\tau_0}]$ where the precessional frequency and the relaxation time are given by $\omega_0 = \gamma\sqrt{(H_{//} + H_{ext})(H_\perp + H_{ext})}$ and $1/\tau_0 = 0.5\alpha\gamma(H_\perp + H_{//} + 2H_{ext})$ respectively. The complex constants A and B are determined by equation (4) and the values of $m_y$ and $m_z$ at t=0. From the above expressions, we see that with given values of $m_y$ and $m_z$ at t=0, the relaxation term is such that the average values of $m_y$ and $m_z$ go to 0 as $t \to \pm\infty$

Let's now see how the feedback circuit is going to modify above equations: The resistance of the MTJ depends on the magnetization direction of the free layer as $R(t) = R_P + (\Delta R/2)(1-m_y(t))$ where $R_P$ is the resistance in parallel state and $\Delta R$ is the difference between the resistances in anti-parallel and parallel states. Thus when dc current is passed through MTJ, a fluctuating voltage of $-I_{dc} m_y(t) \Delta R/2$ appears across it. (Note: The free layer magnetization undergoes thermal fluctuation which is described by the random magnetic field in equation (1)). The fluctuating ac voltage across the MTJ, drives ac current through the feed-back wire. We now assume that the feed-back circuit is terminated into a resistance $R_T$ which is same as the characteristic impedance of the feed-back circuit. Thus ac current flowing through the feed-back wire at position below the MTJ is given by:
$I_{ac}(t) = -0.5 I_{dc} m_y(t-\Delta t) \Delta R /(R_T + R_{MTJ})$ where $R_{MTJ}$ is the average resistance of the MTJ. The ac current flowing below the MTJ depends on the value of $m_y$ at time ($t$-$\Delta t$) due to the delay element and cables connected in the circuit as shown in Fig.1. The resulting ac magnetic field which acts on the free layer is given by $h_{ac} \approx I_{ac}/2w$ [11], where $w$ is the width of the feedback wire. We can write this magnetic field as $h_{ac}(t) = -f I_{dc} m_y(t-\Delta t) \hat{y}$ where the factor $f$ is defined as $f = \Delta R /[4w(R_T + R_{MTJ})]$

We thus have an additional magnetic field along the y-direction due to the feedback circuit which should be included in the LLG equation. With this additional magnetic field, the equations (4a) and (4b) are modified as:

$$\frac{d}{dt}\langle m_y \rangle = -\gamma(H_\perp + H_{ext})\langle m_z \rangle \mp \alpha\gamma(H_{//} + H_{ext})\langle m_y \rangle \pm \alpha\gamma f I_{dc}\langle m_y(t-\Delta t)\rangle \text{ (5a)}$$

$$\frac{d}{dt}\langle m_z \rangle = \gamma(H_{//} + H_{ext})\langle m_y \rangle \mp \alpha\gamma(H_\perp + H_{ext})\langle m_z \rangle - \gamma f I_{dc}\langle m_y(t-\Delta t)\rangle \text{ (5b)}$$

It should be noted that to solve these equations, we need the average value of $m_y$ at time *(t-Δt)*. We now assume solutions of the same form as the previous case, but with different precessional frequency ($\omega$) and relaxation time ($\tau$) for average $m_y$ and $m_z$ i.e.
$\langle m_y \rangle = \text{Re}[A e^{-i\omega t} e^{-|t|/\tau}]$ and $\langle m_z \rangle = \text{Re}[B e^{-i\omega t} e^{-|t|/\tau}]$
We can thus write above two equations (for forward direction of time) as:

$$-(i\omega + \frac{1}{\tau})A = -\gamma(H_\perp + H_{ext})B - \alpha\gamma(H_{//} + H_{ext})A + \alpha\gamma f I_{dc} A e^{i\omega_1 \Delta t} e^{-\Delta t/\tau_1}$$

$$= -\gamma(H_\perp + H_{ext})B - \alpha\gamma(\tilde{H}_{//} + H_{ext})A \text{ (6a)}$$

$$-(i\omega + \frac{1}{\tau})B = \gamma(H_{//} + H_{ext})A - \alpha\gamma(H_\perp + H_{ext})B - \gamma f I_{dc} A e^{i\omega_1 \Delta t} e^{-\Delta t/\tau_1}$$

$$= \gamma(\tilde{H}_{//} + H_{ext})A - \alpha\gamma(H_\perp + H_{ext})B \text{ (6b)}$$

where $\tilde{H}_{//} = H_{//} - f I_{dc} e^{i\omega \Delta t} e^{-\Delta t/\tau}$ is the modified in-plane anisotropy field. This field in general is complex and can be written as $\tilde{H}_{//} = (H_{//} + \delta\tilde{H}_{//}^R) + i\tilde{H}_{//}^I$ where the superscripts *R* and *I* denote the real and imaginary parts respectively. An approximate expression for precessional frequency and relaxation time obtained from equation (6) is given below:

$$\omega \approx \omega_0(1 + \frac{\delta\tilde{H}_{//}^R}{2(H_{//} + H_{ext})} + \frac{\alpha\gamma \tilde{H}_{//}^I}{2\omega_0}) \text{ (7a)}$$

$$\frac{1}{\tau} \approx \frac{1}{\tau_0}[1 - \frac{\tilde{H}_{//}^I}{\alpha(H_\perp + H_{//} + 2H_{ext})}\sqrt{\frac{(H_\perp + H_{ext})}{(H_{//} + H_{ext})}} + \frac{\delta\tilde{H}_{//}^R}{(H_\perp + H_{//} + 2H_{ext})}] \text{ (7b)}$$

where $\omega_0$ and $\tau_0$ are the precession frequency and relaxation time without the feedback circuit. We can see from above equation that the change in real part of in-plane anisotropy field mainly changes the precession frequency, while imaginary part mainly changes the relaxation time.

Let's see what happens if we choose a $\Delta t$ such that $\omega \Delta t = \pi/2$. (The value of $\Delta t$ can be adjusted by the delay element shown in Fig.1). We further assume that $\Delta t/\tau \ll 1$. Thus the effective in-plane anisotropy field is given by $\tilde{H}_{//} = H_{//} - i f I_{dc}$. From equation (7), we find that $\omega \approx \omega_0$ and the relaxation time is modified as:

$$\frac{1}{\tau} \approx \frac{1}{\tau_0}[1 + \frac{f I_{dc}}{\alpha(H_\perp + H_{//} + 2H_{ext})}\sqrt{\frac{(H_\perp + H_{ext})}{(H_{//} + H_{ext})}}] \quad (8)$$

We thus see that positive value of the dc current increases the damping while negative value decreases the damping. If we choose $\Delta t$ such that $\omega \Delta t = 3\pi/2$, the modified $H_{//}$ would be $\tilde{H}_{//} = H_{//} + i f I_{dc}$, which means that positive current would decrease damping and negative current would increase it. If we choose $\Delta t = 0$ or $\omega \Delta t = \pi$, the modified in-plane anisotropy field is real, and for small dc current the dominant effect is change in the precessional frequency. For a general value of $\Delta t$, the change in real as well as imaginary part of the in-plane anisotropy field modifies both the precession frequency and relaxation time.

For making oscillator, we are interested in changing the damping. We therefore choose a $\Delta t$ such that $\omega_0 \Delta t = (2n+1)\pi/2$, where n is non-negative integer. However large value of *n* implies large value of $\Delta t$, and we should not neglect the factor of $exp(-\Delta t/\tau)$ in the effective coercive field: $\tilde{H}_{//} = H_{//} - f I_{dc} e^{i\omega \Delta t} e^{-\Delta t/\tau}$. This factor decreases the effect of feedback circuit. Thus $\omega \Delta t = \pi/2$ is the best choice for changing damping. From equation (8), we see that for certain large negative dc current (critical current), $1/\tau$ becomes zero (i.e. damping of magnetization becomes zero), implying the onset of oscillations. If the out-of-plane anisotropy field is very large, i.e. $H_\perp \gg H_{//}, H_{ext}$, the expression for the critical current obtained from equation (8) comes out to be

$$I_c \approx -\alpha\sqrt{H_\perp(H_{//} + H_{ext})}/f \approx -\alpha \omega_0 / f\gamma \quad (9)$$

The above expression actually corresponds to critical current at zero temperature due to the following reason. As we pass negative current (below the critical current) through MTJ, the damping decreases. This in turn implies that thermal fluctuations increase.[12,13] This means that the assumption of small fluctuations required to obtain linearized LLG equation would no longer hold for large negative currents. The fluctuations for a given value of dc current (less than critical current) depend on the temperature. Lower the temperature, smaller are the fluctuations. This means that the assumption of small fluctuations is valid till the critical current in the limit of zero temperature.

It should be noted that in the above analysis we have shown that the equilibrium position $m_x=1$ is unstable for dc current more than the critical current. By carrying out similar analysis we can show that even $m_x=-1$ state is unstable if the external magnetic field is zero. Thus the free layer magnetization is driven into spontaneous oscillations if the dc current exceeds the critical value. Now consider the case where external magnetic field is applied along the easy axis such that $H_{ext} < H_{//}$. Now the resonant frequency of oscillation is different for $m_x=1$ and $m_x=-1$ states. Since the critical current depends on the oscillation frequency as well as on the factor $\omega \Delta t$, it may happen that one of the two states is stable and other unstable. If we apply field more than $H_{//}$, then for current more than critical current, none of the two states is stable and free layer is driven into oscillations.

Let's now estimate the value of the critical current for a typical MTJ. Let's consider a MgO based MTJ [14, 15] with *$R_P=100$ Ω, $\Delta R=100$ Ω, $R_T=50$ Ω, $\alpha=10^{-2}$, $H_\perp=10^4$ Oe, $H_{//}$* $=100$ Oe, $M_s=800$ emu/cc, $\gamma=2.21 \times 10^5$ m/(A.s) and free layer dimensions of 300 nm x100 nm x4 nm. We can take the width of the feedback wire as the width of the free layer, i.e. *w=100 nm*. With these parameters, the resonant frequency and critical current comes out be $f_0=2.8$GHz and $I_c= -0.64$ mA respectively. It should be noted that the value of critical current is independent of the length and thickness of the free layer, as long as we keep the resonance frequency the same.

We have numerically integrated the differential equation (1) including the feedback term, using Stratonovich calculus and stochastic Heun scheme.[9] The spectral density of $m_y$ (defined as fourier transform of the correlation function) as a function of frequency for above parameters of the free layer is shown in Fig. 2. The green, black and red curves show the spectral density obtained for -0.2 mA, 0 mA and 0.2 mA dc current respectively. The value of *Δt* was chosen such that $\omega \Delta t=\pi/2$. We can clearly see from this figure that for positive current, the spectral density is reduced, while for negative current it is enhanced. Thus the damping is enhanced for positive current, and reduced for negative current as discussed earlier. The blue curve shows the large spectral density obtained for -1 mA, which is more than the critical current ($\omega \Delta t=0.6\pi$ was used for this calculation). The precession frequency obtained is different from the low current case. It should be noted that for large current, the predictions of linearized LLG equation are not valid. We have neglected the effect of spin transfer torque in this calculation. This can be a reasonable approximation if the total magnetic moment of the free layer is large.

The dependence of critical voltage on the width of the free layer is shown in Fig. 3. In this plot, it is assumed that the length of the free layer is 4 times the width, and the resonance frequency is held constant (by adjusting external magnetic field) as we change the width. The plot is obtained by assuming a MTJ with resistance-area product=3 $\Omega(\mu m)^2$, $\alpha=10^{-2}$, $\gamma=2.21 \times 10^5$ m/(A.s), 100% magneto-resistance, $R_T=50$ $\Omega$ and resonance frequency of 2.8 GHz. From Fig. 2, the optimum value of the width is about 100 nm. A better scaling behaviour can be obtained by taking the resistance of the feedback wire as load, instead of the *50 Ω* load as shown in Fig. 1.

The important aspect of the feedback oscillator scheme is that the critical current is independent of the thickness and magnetization of the free layer. (It should be noted that the spin-transfer torque decreases with increasing thickness and magnetization of free layer). For spintronic feedback oscillators we should choose a free layer with large magnetization and thickness. This would increase the total magnetic moment of the free layer and decrease the strength of the thermal fluctuations D (See equation (2)). As a result the line width of the oscillation would be smaller. We can also combine the spin transfer torque effect and feedback circuit in the same device for designing oscillators.

In summary, we have presented a design of spintronics oscillator. A MTJ can be driven into spontaneous oscillations with dc current and a magnetic field feedback circuit.

Figures:

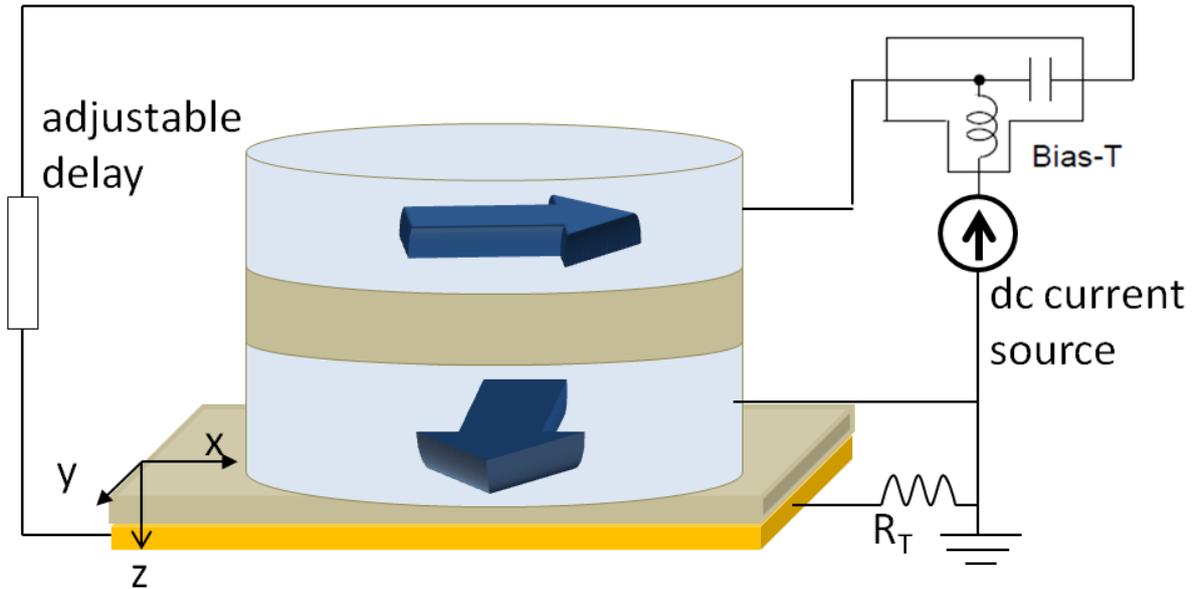

Fig1: Schematic diagram of the feedback oscillator circuit. The top layer of the MTJ pillar shows the free layer, middle layer shows the tunnelling barrier and bottom layer shows the pinned layer. The MTJ rests on the top of a waveguide which is electrically insulated from the MTJ. The wave guide is terminated into a resistance $R_T$ as shown. A fluctuating voltage is produced across the MTJ by the dc current and thermal fluctuations of the free layer. This drives a fluctuating current through the bottom waveguide, and exerts a fluctuating magnetic field on the free layer. The phase of the magnetic field with respect to the free layer oscillation can be adjusted by the adjustable delay. By choosing a suitable value of the phase, the free layer fluctuations can be amplified.

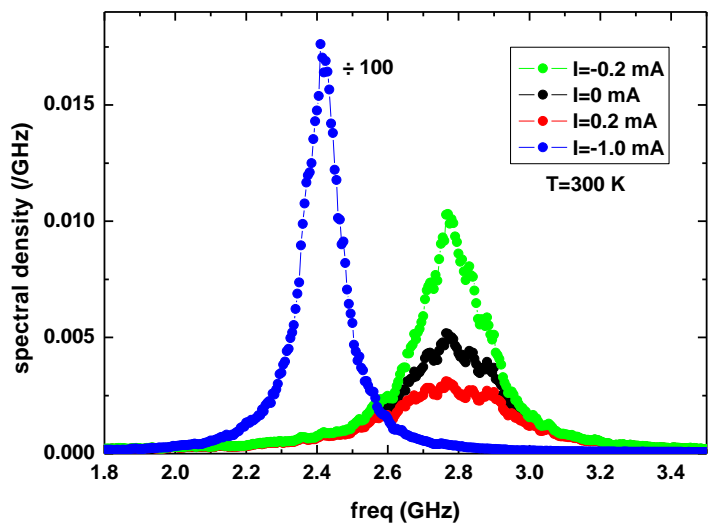

Fig2: The spectral density of spectral density of $m_y$ as a function of frequency at 300K numerically calculated from equation1 including the feedback term. The green, black and red curves show the spectral density with -0.2 mA, 0 mA and 0.2 mA dc current respectively. The damping is enhanced for positive current and reduced for negative current. The blue curve shows the large spectral density obtained with -1 mA current. The spectral density corresponding to the blue curve is divided by 100 in the figure.

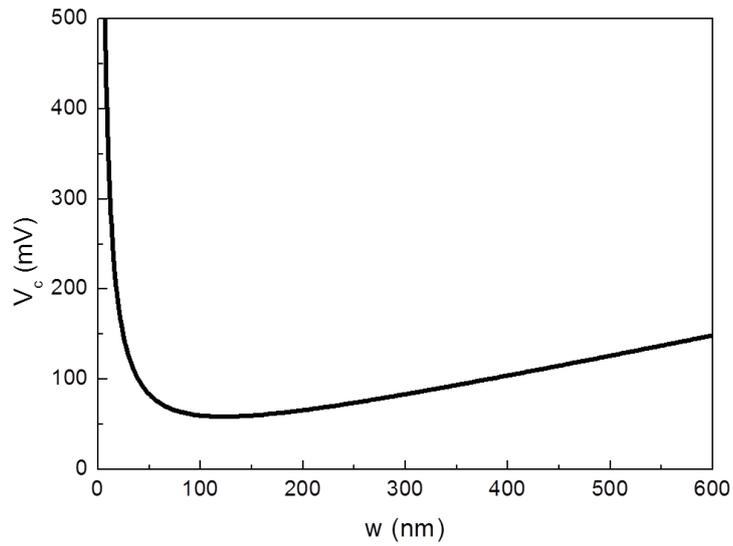

Fig3: The critical voltage required for spontaneous oscillations as a function of the width of the free layer. The width of the feed-back waveguide is assumed to be equal to the width of the free layer. The length of the free layer is taken to be 4 times the width. Other parameters of the MTJ are mentioned in the text.